\documentclass[conference]{IEEEtran}
\IEEEoverridecommandlockouts
\usepackage{cite}
\usepackage{url}
\makeatletter
\newcommand*{\rom}[1]{\expandafter\@slowromancap\romannumeral #1@}
\makeatother
\usepackage{amsmath,amssymb,amsfonts}
\usepackage{algorithmic}
\usepackage{graphicx}
\usepackage{textcomp}
\usepackage{xcolor}
\def\BibTeX{{\rm B\kern-.05em{\sc i\kern-.025em b}\kern-.08em
    T\kern-.1667em\lower.7ex\hbox{E}\kern-.125emX}}
\begin{document}

\title{Generating EEG features from Acoustic features\\
{%\footnotesize \textsuperscript{*}Note: Sub-titles are not captured in Xplore and
}
\thanks{*Equal author contribution}
}

\author{\IEEEauthorblockN{Gautam Krishna}
\IEEEauthorblockA{\textit{Brain Machine Interface Lab} \\
\textit{The University of Texas at Austin}\\
Austin, Texas \\
}
\and
\IEEEauthorblockN{Co Tran}
\IEEEauthorblockA{\textit{Brain Machine Interface Lab} \\
\textit{The University of Texas at Austin}\\
Austin, Texas \\
}
\and
\IEEEauthorblockN{Mason Carnahan*}
\IEEEauthorblockA{\textit{Brain Machine Interface Lab} \\
\textit{The University of Texas at Austin}\\
Austin, Texas \\
}
\and
\IEEEauthorblockN{Yan Han*}
\IEEEauthorblockA{\textit{Brain Machine Interface Lab} \\
\textit{The University of Texas at Austin}\\
Austin, Texas \\
}
\and
\IEEEauthorblockN{Ahmed H Tewfik}
\IEEEauthorblockA{\textit{Brain Machine Interface Lab} \\
\textit{The University of Texas at Austin}\\
Austin, Texas  \\
}
}

\maketitle

\begin{abstract}
In this paper we demonstrate predicting electroencephalography (EEG) features from acoustic features using recurrent neural network (RNN) based regression model and generative adversarial network (GAN). We predict various types of EEG features from acoustic features. We compare our results with the previously studied problem on speech synthesis using EEG and our results demonstrate that EEG features can be generated from acoustic features with lower root mean square error (RMSE), normalized RMSE values compared to generating acoustic features from EEG features (ie: speech synthesis using EEG) when tested using the same data sets. 
\end{abstract}


\section{Introduction}

Electroencephalography (EEG) is a non invasive way of measuring electrical activity of human brain. EEG sensors are placed on the scalp of a subject to obtain the EEG recordings. The references \cite{krishna20,krishna2019state,krishna2019speech} demonstrate that EEG features can be used to perform isolated and continuous speech recognition where EEG signals recorded while subjects were speaking or listening, are translated to text using automatic speech recognition (ASR) models. In \cite{anumanchipalli2019speech} authors demonstrated synthesizing speech from invasive electrocorticography (ECoG) signals using deep learning models. Similarly in \cite{krishna2020synthesis,krishna2019state} authors demonstrated synthesizing speech from EEG signals using deep learning models. 
In \cite{krishna2020synthesis,krishna2019state} authors demonstrated results using different types of EEG feature sets. Speech synthesis and speech recognition using EEG features might help people with speaking disabilities or people who are not able to speak with speech restoration. 

In this paper we are interested in investigating whether it is possible to predict EEG features from acoustic features. This problem can be formulated as an inverse problem of EEG based speech synthesis. In EEG based speech synthesis, acoustic features are predicted from EEG features as demonstrated by the work explained in references \cite{krishna2020synthesis,krishna2019state}. Predicting EEG features or signatures from unique acoustic patters might help in better understanding of how human brain process speech perception and production. Recording EEG signals in a laboratory is a time consuming and expensive process which requires the use of specialized EEG sensors and amplifiers, thus having a computer model which can generate EEG features from acoustic features might also help with speeding up the EEG data collection process as it is much easier to record speech or audio signal, especially for the task of collecting EEG data for performing speech recognition experiments.

In \cite{esteban2017real} authors demonstrated medical time series generation using conditional generative adversarial networks \cite{mirza2014conditional} for toy data sets. Other related work include the reference \cite{hartmann2018eeg} where authors demonstrated generating EEG for motor task using wasserstein generative adversarial networks \cite{arjovsky2017wasserstein}. Similarly in \cite{aznan2019simulating} authors generate synthetic EEG using various generative models for the task of steady state visual evoked potential classification. In \cite{luo2018eeg} authors demonstrated EEG data augmentation for the task of emotion recognition. Our work focuses only on generating EEG features from acoustic features.

We first performed experiments using the model used by authors in \cite{krishna2020synthesis} and then we tried performing experiments using generative adversarial networks (GAN) \cite{goodfellow2014generative}. In this work we predict various EEG feature sets introduced by authors in \cite{krishna2019state} from acoustic features extracted from the speech of the subjects as well as from acoustic features extracted from the utterances that the subjects were listening. 

Our results demonstrate that predicting EEG features from acoustic features seem to be easier compared to predicting acoustic features from EEG features as the root mean square error (RMSE) values during test time were much lower for predicting EEG features from acoustic features compared to it's inverse problem when tested using the same data sets. To the best of our knowledge this is the first time predicting EEG features from acoustic features is demonstrated using deep learning models.  

\section{Regression and GAN model}
 
The regression model we used in this work was very similar to the ones used by the authors in \cite{krishna2020synthesis}. We used the exact training parameters used by authors in \cite{krishna2020synthesis} for setting values for batch size, number of training epochs, learning rate etc. In \cite{krishna2020synthesis} authors used only gated recurrent unit (GRU) \cite{chung2014empirical} layers in their model but in this work we also tried performing experiments using Bi directional GRU layers where a forward GRU and backward GRU layer outputs are concatenated to produce the output of the bi directional GRU layer. The architecture of our regression model is described in Figure 1. The model takes acoustic features or mel-frequency cepstral coefficients (MFCC) of dimension 13 as input and outputs EEG features of a specific dimension at every time step. The dimension of the EEG features outputted depends on the EEG feature set used during training, as each EEG feature set had a different dimension value. The time distributed dense layer in the model has number of hidden units equal to the dimension of the EEG feature set used. The mean squared error (MSE) function was used as the regression loss function for the model. The Figure 4 shows the training convergence for the regression model when Bi directional GRU layers were used. There were two Bi-GRU layers with 256 and 128 hidden units respectively.  

Generative adversarial network (GAN)  \cite{goodfellow2014generative} consists of two networks namely the generator model and the discriminator model which are trained simultaneously. The generator model learns to generate data from a latent space and the discriminator model evaluates whether the data generated by the generator is fake or is from true data distribution. The training objective of the generator is to fool the discriminator. The main motivation behind trying to perform experiments using GAN was in the case of GAN the loss function is learned where as in regression a fixed loss function (MSE) is used. However GAN models are extremely difficult to train. 

Our generator model, as shown in Figure 2, consists of two layers of Bi-GRU with 256, 128 hidden units respectively  in each layer followed by a time distributed dense layer with hidden units equal to the dimension of EEG feature set. During training, real MFCC features with dimension 13 from training set are fed into the generator model and the generator outputs a vector of dimension equal to EEG feature set dimension, which can be considered as fake EEG. 

The discriminator model, as described in Figure 3, consists of two single layered Bi-GRU with 256, 128 hidden units connected in parallel. At each training step a pair of inputs are fed into the discriminator. The discriminator takes (real MFCC features, fake EEG) and (real MFCC features, real EEG) pairs.  The outputs of the two parallel Bi-GRU's are concatenated and then fed to a GRU layer with 128 hidden units. The last time step output of the GRU layer is fed into the dense layer with sigmoid activation function. 

In order to define the loss functions for both our generator and discriminator model let us first define few terms. Let $P_{s_f}$ be the sigmoid output of the discriminator for (real MFCC features, fake EEG) input pair and let $P_{s_e}$ be the sigmoid output of the discriminator for (real MFCC features, real EEG) input pair during training time. Then we can define the loss function of generator as $-\log (P_{s_f}) + (real EEG - fake EEG)^2 *0.5 $ and loss function of discriminator as $-\log (P_{s_e}) - \log(1-P_{s_f})$. The weights of Bi-GRU layers in the generator model were initialized with weights of the regression model for easier training. During test time, the trained generator model takes acoustic features or MFCC from test set as input and produces EEG features as output. 

The Figure 6 shows the generator model training loss and Figure 7 shows the discriminator model training loss.  The GAN model was trained for 200 epochs using adam optimizer with a batch size of 32.

\begin{figure}[h]
\begin{center}
\includegraphics[height=8.5cm, width=\linewidth,trim={0.1cm 0.1cm 0.1cm 0.1cm}]{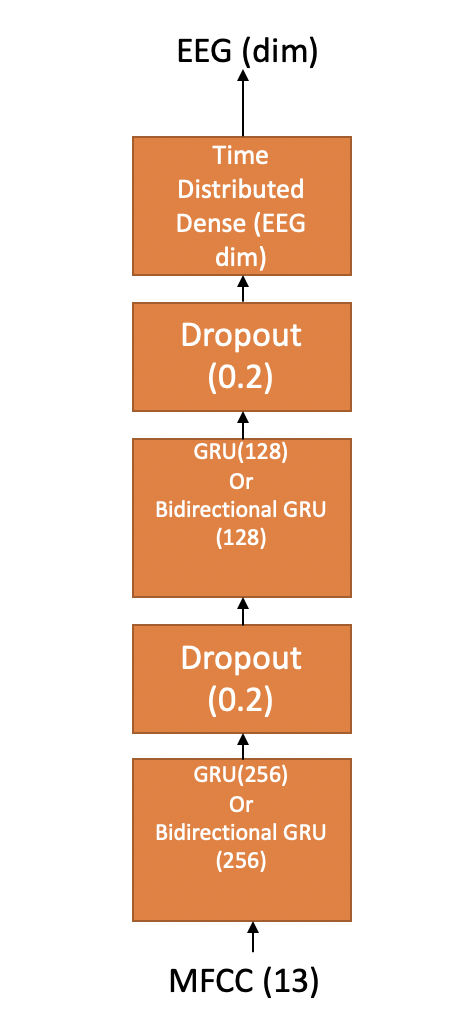}
\caption{Regression Model} 
\label{1vsall}
\end{center}
\end{figure}

\begin{figure}[h]
\begin{center}
\includegraphics[height=8.5cm, width=\linewidth,trim={0.1cm 0.1cm 0.1cm 0.1cm}]{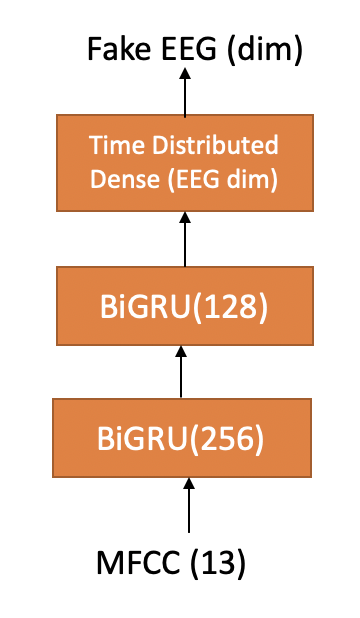}
\caption{Generator in GAN Model} 
\label{1vsall}
\end{center}
\end{figure}

\begin{figure}[h]
\begin{center}
\includegraphics[height=8.5cm, width=\linewidth,trim={0.1cm 0.1cm 0.1cm 0.1cm}]{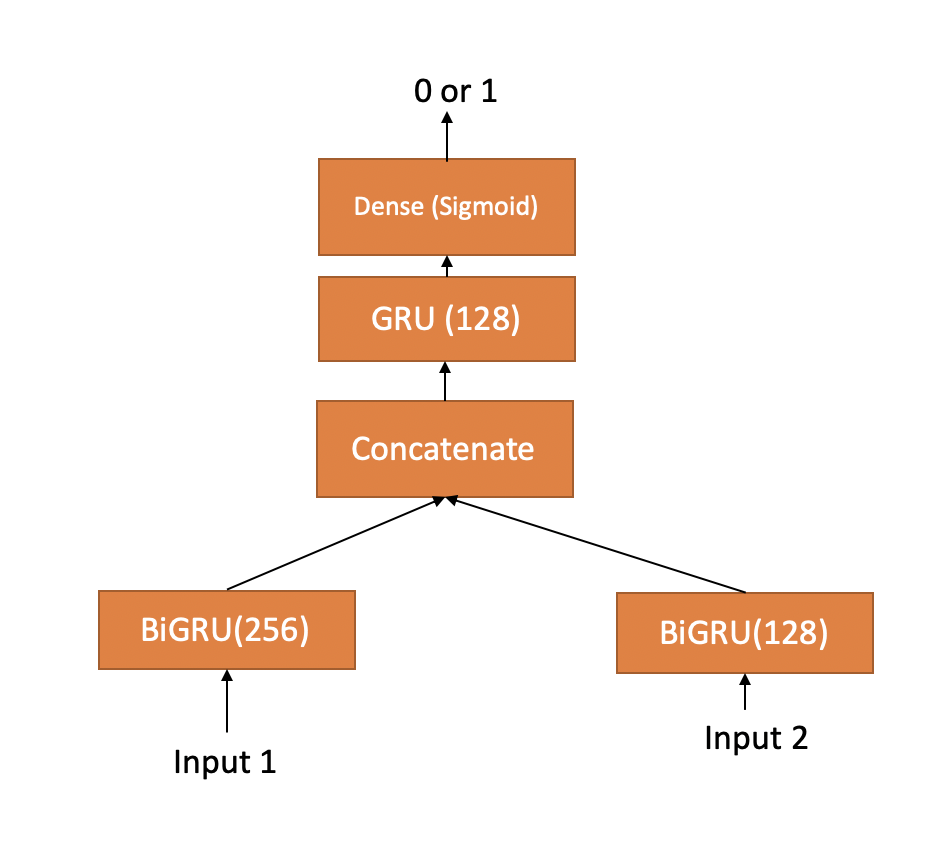}
\caption{Discriminator in GAN Model} 
\label{1vsall}
\end{center}
\end{figure}

\begin{figure}[h]
\begin{center}
\includegraphics[height=5cm, width=0.4
\textwidth,trim={0.1cm 0.1cm 0.1cm 0.1cm},clip]{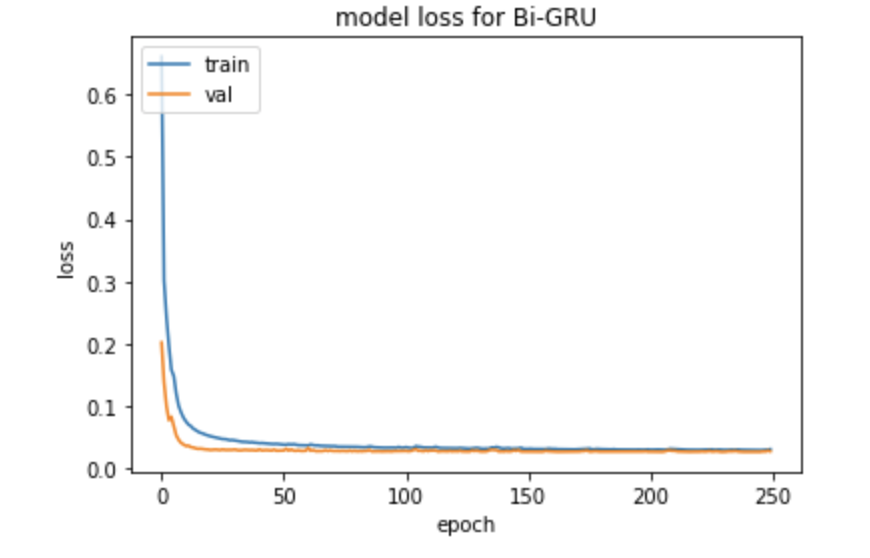}
\caption{Bi-GRU training loss convergence} 
\label{1vsall}
\end{center}
\end{figure}

\begin{figure}[h]
\begin{center}
\includegraphics[height=3cm,width=0.25\textwidth,trim={1cm 1cm 1cm 0.1cm},clip]{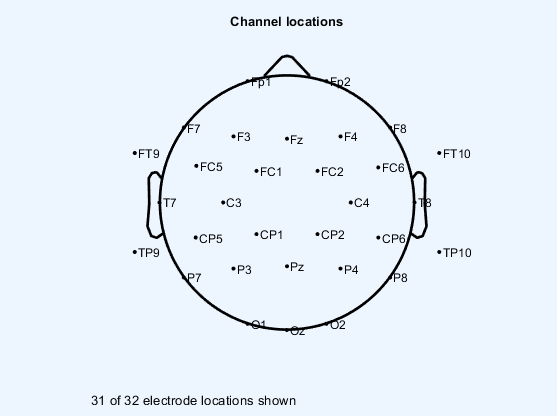}
\caption{EEG channel locations for the cap used in our experiments} 
\label{1vsall}
\end{center}
\end{figure}

\begin{figure}[h]
\begin{center}
\includegraphics[height=5cm, width=0.4
\textwidth,trim={0.1cm 0.1cm 0.1cm 0.1cm},clip]{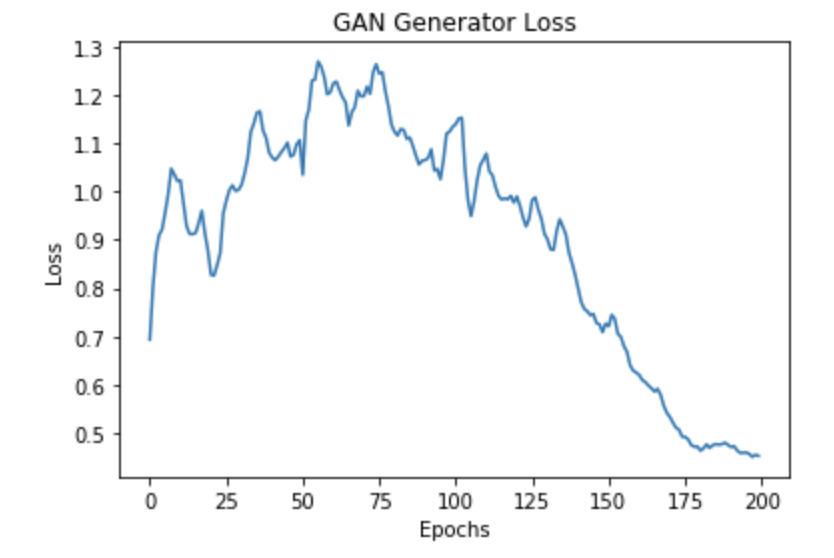}
\caption{Generator training loss} 
\label{1vsall}
\end{center}
\end{figure}

\begin{figure}[h]
\begin{center}
\includegraphics[height=5cm, width=0.4
\textwidth,trim={0.1cm 0.1cm 0.1cm 0.1cm},clip]{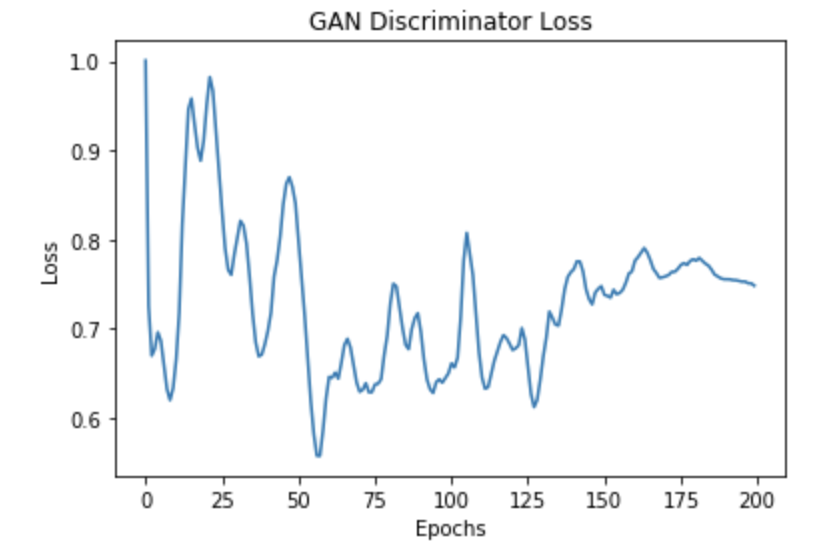}
\caption{Discriminator training loss} 
\label{1vsall}
\end{center}
\end{figure}

\section{Data Sets used for performing experiments}

We used the data set used by authors in \cite{krishna2020synthesis} for performing experiments. The data set contains the simultaneous speech and EEG recording for four subjects. For each subject we used 80\% of the data as the training set, 10\% as validation set and remaining 10\% as test set. This was the main data set used in this work for comparisons. More details of the data set is covered in \cite{krishna2020synthesis}. We will refer this data set as data set A in this paper. 

We also performed some experiments using data set B used by authors in \cite{krishna2019state}. For this data set we didn't perform experiments for each subject instead we used 80\% of the total data as training set, 10\% as validation set and remaining 10\% as test set. More details of the data set is covered in \cite{krishna2019state}. We will refer this data set as data set    B in this paper.
The train-test split was done randomly. 

The EEG data used in these data sets were recorded using wet EEG electrodes. In total 32 EEG sensors were used including one electrode as ground as shown in Figure 5. The Brain Product's ActiChamp EEG amplifier was used in the experiments to collect data.

\section{EEG feature extraction details}

We followed the same preprocessing methods used by authors in \cite{krishna20,krishna2019state,krishna2019speech,krishna2020synthesis} for preprocessing EEG and speech data.

EEG signals were sampled at 1000Hz and a fourth order IIR band pass filter with cut off frequencies 0.1Hz and 70Hz was applied. A notch filter with cut off frequency 60 Hz was used to remove the power line noise.
The EEGlab's \cite{delorme2004eeglab} Independent component analysis (ICA) toolbox was used to remove biological signal artifacts like electrocardiography (ECG), electromyography (EMG), electrooculography (EOG) etc from the EEG signals. 
We then extracted the three EEG feature sets explained by authors in \cite{krishna2019state}. The details of each EEG feature set are covered in \cite{krishna2019state}. Each EEG feature set was extracted at a sampling frequency of 100 Hz for each EEG channel \cite{krishna2019speech}. 

The recorded speech signal was sampled at 16KHz frequency. We extracted mel-frequency cepstral coefficients (MFCC) of dimension 13 as features for speech signal. The MFCC features were also sampled at 100Hz same as the sampling frequency of EEG features.

\section{EEG Feature Dimension Reduction Algorithm Details}

By following the dimension reduction methods used by authors in \cite{krishna2019state} we reduced EEG feature set 1 to a dimension of 30, EEG feature set 2 was reduced to a dimension of 50 using kernel principal component analysis (KPCA) \cite{mika1999kernel} and EEG feature set 3 was kept at original dimension of 93. More details of explained variance plots used to identify the right feature dimensions are covered in \cite{krishna2019state}.

\section{Results}

We computed root mean squared error (RMSE) between the predicted EEG during test time and ground truth EEG from test set as the major performance metric to evaluate the performance of the models during test time for Data set A per subject and for Data set B.  

Tables \rom{1},\rom{2},\rom{3} and \rom{4} shows the results obtained for predicting various listen EEG feature sets from acoustic features for the four subjects belonging to Data set A using GRU and Bi-GRU regression models during test time. Listen EEG refers to the EEG signals recorded while subjects were listening to the utterances. 

Tables \rom{5},\rom{6},\rom{7} and \rom{8} shows the results obtained for predicting various spoken EEG feature sets from acoustic features for the four subjects belonging to Data set A using GRU and Bi-GRU regression models during test time. Spoken EEG refers to the EEG signals recorded while subjects were speaking out loud the utterances. 

We observed that RMSE values were comparable for different EEG feature sets and both GRU, Bi-GRU layers demonstrated similar results. We also computed normalized RMSE as defined by authors in \cite{krishna2020synthesis} and observed an average normalized RMSE of \textbf{0.00068} for spoken condition for each subject and an average normalized RMSE of \textbf{0.0006} for listen condition for each subject belonging to Data set A. Our results demonstrate that the test time average RMSE and normalized RMSE values were significantly lower than values obtained by authors in \cite{krishna2020synthesis} where they were predicting acoustic features from EEG features. These results demonstrate it is easier for a deep model to learn the mapping from acoustic features to EEG features rather than trying to learn the mapping from EEG to acoustic features. 

When we performed experiments using GAN model on data set A for each subject during test time we observed an average RMSE of \textbf{0.36} for spoken, listen condition for each EEG feature set. Thus the GRU and Bi-GRU regression models outperformed GAN for predicting EEG features from acoustic features. Even though we added regularization terms to the loss function of the generator in our GAN model, it still didn't help to outperform regression models. Hypothetically GAN should have demonstrated better results than regression model as GAN also learns the loss function. Our results demonstrate the extreme difficulty of training GAN for sequence generation task. The results presented by authors in \cite{krishna2019state} also demonstrate that RNN models outperformed GAN for the task of predicting acoustic features from EEG features. 

We performed experiments using GRU regression model for Data set B and observed an average RMSE of \textbf{0.23} for spoken, listen condition for each EEG feature set during test time.  The observed average RMSE was again much lower compared to the RMSE values obtained by authors in \cite{krishna2019state} where they tried predicting acoustic features from EEG features using the same Data set B. 

Another interesting observation we noted was that in case of the test time results demonstrated by authors in \cite{krishna2020synthesis}, the RMSE values for predicting acoustic features from EEG varied among subjects whereas in our results we observed that RMSE values during test time remained almost constant among the four subjects belonging to Data set A indicating our model was able to generalize better for all the four subjects and it also indicates the deep learning model can learn acoustic to EEG mapping easily compared to learning the mapping from EEG to acoustic features.

\begin{table}[!ht]
\centering
\begin{tabular}{|l|l|l|}
\hline
\textbf{\begin{tabular}[c]{@{}l@{}}EEG\\ Feature\\ Set\end{tabular}} & \textbf{\begin{tabular}[c]{@{}l@{}}Average\\ RMSE\\ GRU \\ Model\end{tabular}} & \textbf{\begin{tabular}[c]{@{}l@{}}Average\\ RMSE\\ Bi-GRU\\ Model\end{tabular}} \\ \hline
Set 1                                                                & 0.23                                                                           & 0.23                                                                             \\ \hline
Set 2                                                                & \textbf{0.20}                                                                  & 0.206                                                                            \\ \hline
Set 3                                                                & \textbf{0.19}                                                                  & 0.193                                                                            \\ \hline
\end{tabular}
\caption{Results for predicting \textbf{listen EEG from listen MFCC for subject 1 Data set A}}
\end{table}

\begin{table}[!ht]
\centering
\begin{tabular}{|l|l|l|}
\hline
\textbf{\begin{tabular}[c]{@{}l@{}}EEG\\ Feature\\ Set\end{tabular}} & \textbf{\begin{tabular}[c]{@{}l@{}}Average\\ RMSE\\ GRU \\ Model\end{tabular}} & \textbf{\begin{tabular}[c]{@{}l@{}}Average\\ RMSE\\ Bi-GRU\\ Model\end{tabular}} \\ \hline
Set 1                                                                & 0.22                                                                           & 0.22                                                                             \\ \hline
Set 2                                                                & \textbf{0.20}                                                                  & 0.21                                                                             \\ \hline
Set 3                                                                & 0.19                                                                           & 0.19                                                                             \\ \hline
\end{tabular}
\caption{Results for predicting \textbf{listen EEG from listen MFCC for subject 2 Data set A}}
\end{table}

\begin{table}[!ht]
\centering
\begin{tabular}{|l|l|l|}
\hline
\textbf{\begin{tabular}[c]{@{}l@{}}EEG\\ Feature\\ Set\end{tabular}} & \textbf{\begin{tabular}[c]{@{}l@{}}Average\\ RMSE\\ GRU \\ Model\end{tabular}} & \textbf{\begin{tabular}[c]{@{}l@{}}Average\\ RMSE\\ Bi-GRU\\ Model\end{tabular}} \\ \hline
Set 1                                                                & 0.23                                                                           & 0.23                                                                             \\ \hline
Set 2                                                                & 0.21                                                                           & 0.21                                                                             \\ \hline
Set 3                                                                & 0.19                                                                           & 0.19                                                                             \\ \hline
\end{tabular}
\caption{Results for predicting \textbf{listen EEG from listen MFCC for subject 3 Data set A}}
\end{table}

\begin{table}[!ht]
\centering
\begin{tabular}{|l|l|l|}
\hline
\textbf{\begin{tabular}[c]{@{}l@{}}EEG\\ Feature\\ Set\end{tabular}} & \textbf{\begin{tabular}[c]{@{}l@{}}Average\\ RMSE\\ GRU \\ Model\end{tabular}} & \textbf{\begin{tabular}[c]{@{}l@{}}Average\\ RMSE\\ Bi-GRU\\ Model\end{tabular}} \\ \hline
Set 1                                                                & 0.25                                                                           & 0.25                                                                             \\ \hline
Set 2                                                                & 0.23                                                                           & 0.23                                                                             \\ \hline
Set 3                                                                & 0.21                                                                           & 0.21                                                                             \\ \hline
\end{tabular}
\caption{Results for predicting \textbf{listen EEG from listen MFCC for subject 4 Data set A}}
\end{table}

\begin{table}[!ht]
\centering
\begin{tabular}{|l|l|l|}
\hline
\textbf{\begin{tabular}[c]{@{}l@{}}EEG\\ Feature\\ Set\end{tabular}} & \textbf{\begin{tabular}[c]{@{}l@{}}Average\\ RMSE\\ GRU \\ Model\end{tabular}} & \textbf{\begin{tabular}[c]{@{}l@{}}Average\\ RMSE\\ Bi-GRU\\ Model\end{tabular}} \\ \hline
Set 1                                                                & 0.23                                                                           & 0.23                                                                             \\ \hline
Set 2                                                                & 0.21                                                                           & 0.21                                                                             \\ \hline
Set 3                                                                & 0.20                                                                           & 0.20                                                                             \\ \hline
\end{tabular}
\caption{Results for predicting \textbf{spoken EEG from spoken MFCC for subject 1 Data set A}}
\end{table}

\begin{table}[!ht]
\centering
\begin{tabular}{|l|l|l|}
\hline
\textbf{\begin{tabular}[c]{@{}l@{}}EEG\\ Feature\\ Set\end{tabular}} & \textbf{\begin{tabular}[c]{@{}l@{}}Average\\ RMSE\\ GRU \\ Model\end{tabular}} & \textbf{\begin{tabular}[c]{@{}l@{}}Average\\ RMSE\\ Bi-GRU\\ Model\end{tabular}} \\ \hline
Set 1                                                                & 0.23                                                                           & 0.23                                                                             \\ \hline
Set 2                                                                & 0.22                                                                           & 0.22                                                                             \\ \hline
Set 3                                                                & 0.21                                                                           & 0.21                                                                             \\ \hline
\end{tabular}
\caption{Results for predicting \textbf{spoken EEG from spoken MFCC for subject 2 Data set A}}
\end{table}

\begin{table}[!ht]
\centering
\begin{tabular}{|l|l|l|}
\hline
\textbf{\begin{tabular}[c]{@{}l@{}}EEG\\ Feature\\ Set\end{tabular}} & \textbf{\begin{tabular}[c]{@{}l@{}}Average\\ RMSE\\ GRU \\ Model\end{tabular}} & \textbf{\begin{tabular}[c]{@{}l@{}}Average\\ RMSE\\ Bi-GRU\\ Model\end{tabular}} \\ \hline
Set 1                                                                & 0.24                                                                           & 0.24                                                                             \\ \hline
Set 2                                                                & 0.23                                                                           & 0.23                                                                             \\ \hline
Set 3                                                                & 0.21                                                                           & 0.21                                                                             \\ \hline
\end{tabular}
\caption{Results for predicting \textbf{spoken EEG from spoken MFCC for subject 3 Data set A}}
\end{table}

\begin{table}[!ht]
\centering
\begin{tabular}{|l|l|l|}
\hline
\textbf{\begin{tabular}[c]{@{}l@{}}EEG\\ Feature\\ Set\end{tabular}} & \textbf{\begin{tabular}[c]{@{}l@{}}Average\\ RMSE\\ GRU \\ Model\end{tabular}} & \textbf{\begin{tabular}[c]{@{}l@{}}Average\\ RMSE\\ Bi-GRU\\ Model\end{tabular}} \\ \hline
Set 1                                                                & 0.24                                                                           & 0.24                                                                             \\ \hline
Set 2                                                                & 0.22                                                                           & 0.22                                                                             \\ \hline
Set 3                                                                & 0.21                                                                           & 0.21                                                                             \\ \hline
\end{tabular}
\caption{Results for predicting \textbf{spoken EEG from spoken MFCC for subject 4 Data set A}}
\end{table}

\section{Conclusion and Future work}

In this paper we demonstrated predicting various EEG feature sets from acoustic features with very \textbf{low RMSE} and \textbf{normalized RMSE} values during test time. To the best of our knowledge this is the first time predicting EEG features from acoustic features is demonstrated using deep models. Our results demonstrate it is easier for a deep model to learn the mapping from acoustic to EEG features rather than trying to map the inverse. 

The future work will focus on validating the results on a larger data set with more number of subjects and developing strategies to improve the training of GAN for the task of generating EEG features from acoustic features. 

Our future work will also focus on using these results to better understand the underlying science behind human brain's ability to perform speech perception and production.

\section{Acknowledgement} 
We would like to thank Kerry Loader and Rezwanul Kabir from Dell, Austin, TX for donating us the GPU to train the models used in this work.

\bibliographystyle{IEEEtran}

\bibliography{refs}
\end{document}